\documentclass[fleqn,usenatbib]{mnras}

\usepackage{newtxtext,newtxmath}

\usepackage[T1]{fontenc}
\usepackage{stfloats}
\DeclareRobustCommand{\VAN}[3]{#2}
\let\VANthebibliography\thebibliography
\def\thebibliography{\DeclareRobustCommand{\VAN}[3]{##3}\VANthebibliography}

\usepackage{graphicx}	
\usepackage{amsmath}	
\usepackage{afterpage}
\usepackage{wrapfig}
\usepackage{caption}

\title[CNN-Derived Elemental Abundances]{CNN-Derived Elemental Abundances of LAMOST DR10 Giants: Implications for Galactic Substructures}

\author[Haoyang Liu et al.]{%
  Haoyang Liu\textsuperscript{1}, 
  Cuihua Du\textsuperscript{1}\thanks{Corresponding author: CHD (co-first author);
  ducuihua@ucas.ac.cn}, 
  Mingji Deng\textsuperscript{1}, 
  Jian Zhang\textsuperscript{1}\\
  \textsuperscript{1}College of Astronomy and Space Sciences, 
  University of Chinese Academy of Sciences, 
  Beijing 100049, P.R. China
}

\pubyear{\the\year{}}

\begin{document}
\label{firstpage}
\pagerange{\pageref{firstpage}--\pageref{lastpage}}
\maketitle

\begin{abstract}
Stellar parameters and abundances provide crucial insights into stellar and Galactic evolution studies. In this work, we developed a convolutional neural network (CNN) to estimate stellar parameters: effective temperature ($T_{\text{eff}}$), surface gravity (log $g$) and metallicity (both [Fe/H] and [M/H]) as well as six $\alpha$-elements (C, N, O, Mg, Si, Ca) and [$\alpha$/M]. We selected giant stars with \( 3500 \, \text{K} < T_{\text{eff}} < 5500 \, \text{K} \) and \( 0 \, \text{dex} < \log g < 3.6 \, \text{dex} \) from the LAMOST and APOGEE surveys, while requiring (S/N)$_g$ of the LAMOST low-resolution spectra $>$ 10, which leaves 1,100,858 giant stars. The spectral from LAMOST and the labels from APOGEE for 62,511 common stars were used as our training set. The corresponding test set yields scatters 50 K, 0.06 dex and 0.13 dex for $T_{\text{eff}}$, [Fe/H] and log $g$, respectively. For $\alpha$ elements O, Mg, Si and Ca, the scatters are 0.05 dex, 0.04 dex, 0.03 and 0.04 dex, respectively. For C and N elements, the scatters are 0.07 dex and 0.05 dex. For [$\alpha$/M] and [M/H], the scatters are 0.03 dex and 0.06 dex. The mean absolute error (MAE) of most elements are between 0.02 $-$ 0.04 dex. The predicted abundances were cross-matched with previously identified substructures PG1 and PG2, with their origins subsequently analyzed. Finally, the catalog is available at: \url{https://nadc.china-vo.org/res/r101529/}.

\end{abstract}




\section{Introduction}\label{sec:intro}
Elemental abundances serve as key tracers for the chemical evolution of the Galaxy. The advent of large-scale spectroscopic surveys has greatly advanced our ability to study these abundances, such as the Large Sky Area Multi-Object Fiber Spectroscopic Telescope \citep[LAMOST;][]{Cui2012}, the Apache Point Observatory Galactic Evolution Experiment \citep[APOGEE;][]{Majewski2017}, the RAdial Velocity Experiment \citep[RAVE;][]{Steinmetz2006}, the Sloan Extension for Galactic Understanding and Exploration \citep[SEGUE;][]{Yanny2009}, the Dark Energy Spectroscopic Instrument \citep[DESI;][]{D2016}, the Galactic Archaeology with HERMES \citep[GALAH;][]{GALAH2015}, \textit{Gaia}-ESO \citep{Gaia2012} and upcoming surveys like the Chinese Space Station Telescope (CSST), are expected to provide more stellar spectra.

The application of deep learning in stellar spectra has become increasingly widespread with the growing power of computational resources. \citet{Ness2015} developed a data-driven approach called \texttt{the Cannon}, which maps APOGEE spectra to known stellar labels, yielding results comparable to those from APOGEE pipeline (ASPCAP) labels. Later \citet{Fabbro2018} created a convolutional neural network (CNN) architecture called \texttt{StarNet} trained on APOGEE synthetic spectra, which demonstrated better performance than \texttt{the Cannon}. \texttt{The Payne} \citep{Ting2019} is a fully connected network with two hidden layers that combines information from the available spectral range and fits labels simultaneously using the adopted spectral models. Based on important ingredients from \texttt{the Cannon} and \texttt{the Payne}, \citet{Xiang2019} built a data-driven Payne (DD-Payne) and estimated sixteen abundances for six million stars from LAMOST DR5. \texttt{astroNN} \citep{Leung2019} is also fascinating because it adopts a Monte-Carlo dropout process to obtain the prediction uncertainties. There are also non-neural network method such as the Stellar LAbel Machine (SLAM; Zhang et al.\citeyear{Zhangbo2020}), which employs a non-linear support vector regression method to avoid the ``when-to-stop" problem. This problem refers to how to find the best number of epochs in case of both overfitting and underfitting in neural network (NN). \citet{Xiang2021} proposed a data-driven model based on statistical features and Catboost algorithm (SCDD), achieving relatively good results for LAMOST DR6 spectra.

While the LAMOST Stellar Pipeline provides fundamental stellar parameters, partial [Fe/H] measurements, and limited [$\alpha$/Fe] abundances, these sparse data are insufficient for robust investigation of accretion signatures in the Galactic halo. Detailed $\alpha$-element abundances enable more robust studies of chemical characteristics of accreted dwarf galaxies, including their $\alpha$-knee features. These measurements provide critical constraints on the star formation histories of their progenitors. Furthermore, $\alpha$-element abundances provide crucial insights into the chemical and structural properties of the Galactic disk. Recent studies have revealed that the high-$\alpha$ disk population does not strictly correspond to the conventionally defined thick disk \citep[see Section 3.1 in the review:][]{Hunt2025}. This highlights the value of applying diverse methodologies to LAMOST low-resolution spectra—as demonstrated both in our work and previous studies-to obtain comprehensive abundance measurements for investigating Galactic structures, with particular emphasis on the Galactic halo (as will be discussed later in this section).

These NN methods have also been largely conducted to predict LAMOST stellar labels. \citet{Ho2017} utilized \texttt{the Cannon} and predicted stellar labels for LAMOST 450,000 giants with scatters of 70 K in $T_{\text{eff}}$, 0.1 dex in both log $g$ and [Fe/H], 0.04 dex in [$\alpha$/M] with $g$-band signal-to-noise ration (S/N)$_{g}$ higher than 50. \citet{zhang2019} applied \texttt{StarNet} to determine stellar parameters and C, N and $\alpha$ abundances for 938,720 Giants from LAMOST DR5, with uncertainties of 45 K for $T_{\text{eff}}$, 0.1 dex for log $g$, 0.05 dex for [M/H], 0.03 dex for [$\alpha$/M], 0.06 dex for [C/M] and 0.07 dex for [N/M]. \citet{Wang2022} used a fully connected NN with three layers to estimate stellar parameters and abundances of 7.10 million stars from LAMOST low-resolution DR8, yielding scatters 111 K for $T_{\text{eff}}$, 0.11 dex for log $g$, 0.05 dex for [Fe/H], 0.09 dex for [N/Fe] and 0.03 dex for [$\alpha$/M] among APOGEE common stars. For medium resolution spectra, \cite{Wang2020} developed a deep learning method called SPCANet to derive 11 elemental abundances for stars from LAMOST-II Medium Resolution Survey. All these methods are related to the label transfer approach. This method leverages APOGEE precise stellar parameters and chemical abundances as reference values for the corresponding LAMOST stars, ensuring accuracy and consistency in our analysis of the LAMOST giant star sample. 
Although APOGEE and LAMOST have disjoint wavelength coverage and spectral resolution, the ``stellar labels" represent the intrinsic properties of the stars. Therefore, proper methods must be employed to ensure that predictions are brought to the same scale and yield unbiased values \citep{Ho2017}. As a result, these aforementioned studies point out that it is a robust way to derive LAMOST stellar labels with an NN method.

The Galactic halo consists of old, metal-poor stellar populations, which are primarily thought to originate from tidally disrupted dwarf galaxies. Studying the Galactic halo could help us better understand the accretion and evolutionary history of the Milky Way (known as ``Galactic archaeology"), which also allows us to test existing theories of galaxy formation. Disrupted dwarf galaxies often leave observable traces in the form of clumps in integrals-of-motion space, owing to their long dynamical relaxation timescales. These clumps, referred to as ``substructures," are considered fossil remnants of the disrupted galaxies. By examining the kinematic and chemical properties of substructures in detail, along with their interconnections, we can construct a more complete understanding of the Milky Way’s accretion history and evolutionary timeline. Giant stars can provide more accurate distance measurements due to the intrinsic brightness, enabling us to study the distant structure of the Galactic halo \citep[e.g.,][]{Yang2019,Han2022}. As a high-precision spectroscopic survey (R$\sim22,500$), APOGEE DR17 \citep{A2022} provides 68,000 targets towards the Galactic halo, while LAMOST offers significantly wider sky coverage. By using the elemental abundances of APOGEE DR17 as reference labels and training on the lower-resolution spectra of LAMOST DR10, we can derive improved abundance measurements that enable more comprehensive studies of the Galactic substructures.

The derived abundances include: light elements (C, N), $\alpha$-elements (O, Mg, Si, Ca, [$\alpha$/M]), and metallicity ([M/H], [Fe/H]). The abundances of $\alpha$-elements serve as excellent tracers of the timescale between Core Collapse supernovae (CCSNe) and the delayed contributions from Type Ia supernovae, thereby helping to reconstruct the star formation history \citep{Timmes1995}. The observed dwarf galaxies typically exhibit lower $\alpha$-element abundances compared to the Milky Way \citep{Hasselquist2021}, suggesting a potentially top-light initial mass function \citep[e.g.,][]{McWilliam2013}. An enhanced $\alpha$-element abundance may also trace the ongoing merger event affecting the Galactic disk \citep[e.g.,][]{Sun2025}. The [C/N] ratio serves as a reliable mass indicator for red giant branch (RGB) stars \citep{Martig2016}, which constitute the majority of giants in our sample. This ratio consequently provides age estimates for these stars, making it a valuable tool for ``Galactic archaeology". We exclude Mn and Al in this work because APOGEE DR17 wavelength coverage poorly samples their strong line features, potentially compromising abundance reliability. An alternative perspective suggests that the selection of accreted stars via the [Mn/Mg]-[Al/Fe] plane may be model-dependent and unreliable \citep{Vasini2024}. We therefore conclude that these predicted elemental abundances are sufficient for investigating the bulk properties of the structures and other halo structures.

In this work, we plan to use CNN to obtain the stellar parameters and abundances of all giant stars  for LAMOST DR10: in section~\ref{data}, we give a brief introduction on the data used and the preprocessing steps applied to the input spectra. In section~\ref{method}, the CNN architecture we use and performances on the training set. In section~\ref{predictions}, we predict the stellar labels of 1,100,858 giants from LAMOST DR10. In section~\ref{summary}, we give a comprehensive summary.

\begin{figure*}
    \centering
    \includegraphics[width=\linewidth]{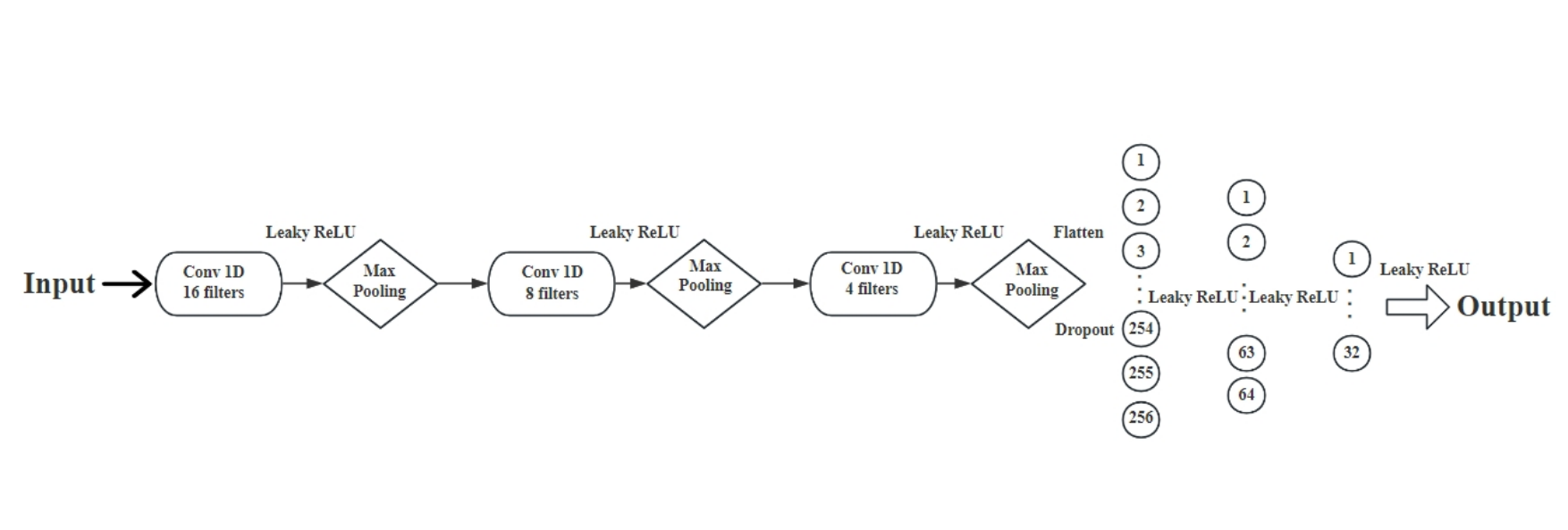}
    \caption{The schematic diagram of CNN architecture used in this work. The model adopts three convolutional layers with 16, 8, and 4 filters of size 10, each followed by a max pool layer of size 2. A dropout layer follows the third convolutional layer, leading to three fully connected layers with 256, 128, and 64 neurons, respectively. Note that LeakyReLU activation function is used in this model.}
    \label{map}
\end{figure*}

\section{Data}\label{data}\label{data}
\subsection{Giants Selection}
The Large sky Area Multi-Object Spectroscopic Telescope (LAMOST) is a low resolution (R $\sim$ 1,800) optical (3650 - 9000 Å) spectroscopic survey. It has a large aperture (effective aperture of 3.6 m$-$4.9 m) and a wide field of view 5°, collecting 4000 fiber spectra simultaneously \citep{Cui2012}. The two-dimension data are processed by LAMOST 2D pipeline for spectral extraction, flat-fielding, background subtraction, wavelength and flux calibration to produce 1D spectra \citep{Luo2015}. The latest publicly available dataset is LAMOST Data Release 10 low resolution (version 2.0), containing 11,441,011 spectra and 11,100,139 stars. The stellar parameters including effective temperatures, surface gravity and metallicity are generated by LAMOST Stellar Pipeline (LASP; Wu et al.\citeyear{Wu2014}). 

The Apache Point Observatory Galactic Evolution Experiment (APOGEE; Majewski et al.\citeyear{Majewski2017}) is a high spectroscopic survey (R $\sim$ 22,500) with high signal-to-noise ratio ($>$ 100) in the near infrared spectral range (1.51 $-$ 1.70 $\mu$m). APOGEE spectra is collected by a Sloan 2.5$-$m telescope with a 3° diameter field of view \citep{Gunn2006}. APOGEE Data Release 17 \citep{A2022} is the latest dataset containing over 650,000 stars with precise elemental abundances estimated by APOGEE Stellar Parameters and Chemical Abundances Pipeline (APSCAP; García Pérez et al.\citeyear{GARC2016}).

To obtain the training and test data sets, we first select giants from both LAMOST and APOGEE by applying cuts: \( 3500 \, \text{K} < T_{\text{eff}} < 5500 \, \text{K} \) and \( 0 \, \text{dex} < \log g < 3.6 \, \text{dex} \) according to LASP and APSCAP estimates, respectively. For LAMOST giants, we additionally require (S/N)$_{g}$ greater than 10 to ensure relatively high-quality spectra, along with the availability of redshift values, as the absence of redshift data indicates that the spectra may be significantly contaminated by cosmic rays. Consequently, we are left with a total of 1,100,858 stars\footnote{The selection criteria could be applied online through the LAMOST LRS Combined Catalogue Query. See \url{https://www.lamost.org/dr10/search}}. Among all these cuts, the signal-to-noise requirement caused the most significant decrease. For APOGEE giants, the corresponding errors for effective temperatures are below 100 K, and for surface gravity, they are below 0.1 dex. We also require that signal-to-noise ratio $>$ 70 and elemental errors $<$ 0.2 dex for reliable labels and this leaves 340,980 stars. We then cross-match giants from the two surveys within 1 acrsecond and 62,511 common giants are left.

\subsection{Data Preprocessing}
For preprocessing the input LAMOST spectra, we normalize the spectra by dividing an error-weighted and broad Gaussian smoothing following the approach in \citet{Ho2017}:
\begin{equation}
    \bar{f}(\lambda_{0}) = \frac{\sum_{i}(f_{i}\sigma_{i}^{-2}\omega_{i}(\lambda_{0}))}{\sum_{i}(\sigma_{i}^{-2}\omega_{i}(\lambda_{0}))}
\end{equation}
where the weight is in a Gaussian form:
\begin{equation}
    \omega_{i}(\lambda_{0}) = e^{-\frac{(\lambda_{0}-\lambda_{i})^2}{L^2}}
\end{equation}
here $f_{i}$, $\sigma_{i}$ are the flux and uncertainty at each pixel respectively, and $L$ is set to be 50 Å because it is much broader than typical atomic lines. In this work, we also remove any pixels with zero inverse variance (means infinity $\sigma$) in each spectrum to avoid bad-quality flux. Note that the normalization described here is not the typical ``continuum normalization", which requires polynomial fitting. Instead, this ``pseudo-normalization" simplifies the modeling procedure by removing overall flux, flux calibration, and large-scale shape variations from the spectra. We do not provide the total number of pixels summed up here because each LAMOST spectrum has a varying number of pixels, and the corresponding values are not identical.

After normalization, we interpolate and extrapolate the spectra into 5,000 bins, spanning the wavelength range from 4,000 to 8,500 Å. This ensures uniform input size and reduces the noise present at both the blue and red ends of the spectra. Finally, the input normalized spectra and stellar labels are standardized to conform to a standard normal distribution, with a mean of 0 and a standard deviation of 1.

\section{METHOD}\label{method}
\subsection{CNN Architecture}
In this work, we apply a CNN architecture using \textit{TensorFlow} framework to train and test the LAMOST spectra of common stars, which contains three convolutional layers and three hidden fully connected layers. Three convolutional layers have 16 filters, 8 filters, and 4 filters, each with a size of 10. The kernel size for all filters follows the recommendations of \citet{Guiglion2020}, indicating that a kernel size between 5 and 20 pixels tends to extract features efficiently. A max pooling layer with a size of 2 follows each convolutional layer to further extract spectroscopic features. The network includes three fully connected layers with 256, 128, and 64 neurons, respectively. The activation function is chosen to be LeakyReLu to avoid ``dying ReLu" problem \citep{relu}:
\begin{equation}
    \text{LeakyReLU}(x) = x \,(x > 0) \: \text{or} \: \gamma x \,(x \leq 0)
\end{equation}
Here, the default value 0.2 is used for $\gamma$ and a dropout is also added after the third convolutional layer in case of overfitting. The whole architecture is shown in Figure~\ref{map}.

\subsection{Training Process}
\begin{figure*}
    \centering
    \includegraphics[width=\linewidth]{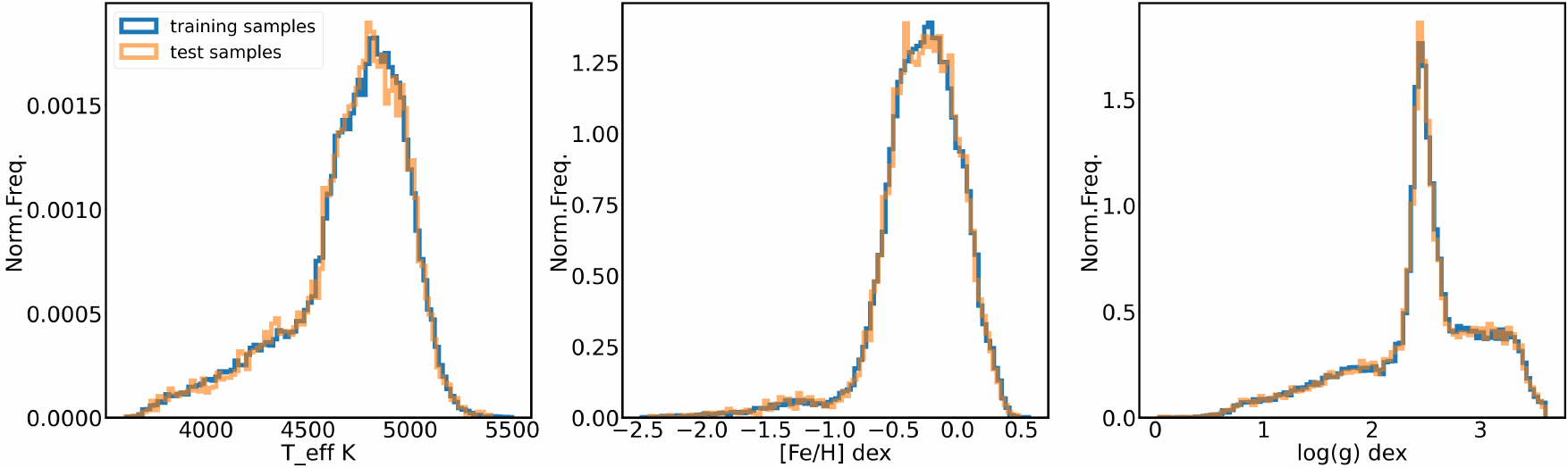}
    \caption{Distributions of training (blue) and test (orange) samples in $T_{\text{eff}}$, [Fe/H] and log $g$. There are 50,008 stars in the training sample and 12,503 stars in the test sample.}
    \label{coverage}
\end{figure*}

In the training process, we use \texttt{train\_test\_split} from \texttt{scikit-learn} python package \citep{scikit-learn} to randomly split the training and test samples with a ratio of 8:2. There are 50,008 stars in the training set and 12,503 stars in the test set after the split (see Figure~\ref{coverage}). Mean absolute error (MAE) is taken as the loss function to evaluate model training process, which is defined as follows:
\begin{equation}
    \text{MAE} = \frac{1}{N}\sum_{i}|y_{i}-y_{i}^{pred}|
\end{equation}
where $N$ stands for the size of samples, $y_{i}$ indicates the real labels and $y_{i}^{pred}$ are the predicted labels. We fine-tune the hyper-parameters according to MAE values, and we find the model performs relatively better if the \texttt{batch\_size} and \texttt{learning\_rate} are set to be 16 and 1e$-$5 respectively. A dropout layer is used to randomly exclude a fraction of neurons to prevent overfitting. \citet{Guiglion2020} found that a dropout fraction between 10 \% and 30 \% does not significantly affect model performance. Therefore, we adopt a dropout fraction of 20 \% consistent with their findings. The Adam optimizer is a combination of two gradient descent methods: Momentum and Root Mean Square Propagation (RMSProp) \citep{Adam}. Given its greater efficiency and lower memory requirements, we choose the Adam optimizer to calculate the weights and biases of the CNN model.

\begin{figure}
    \centering
    \includegraphics[width=\linewidth]{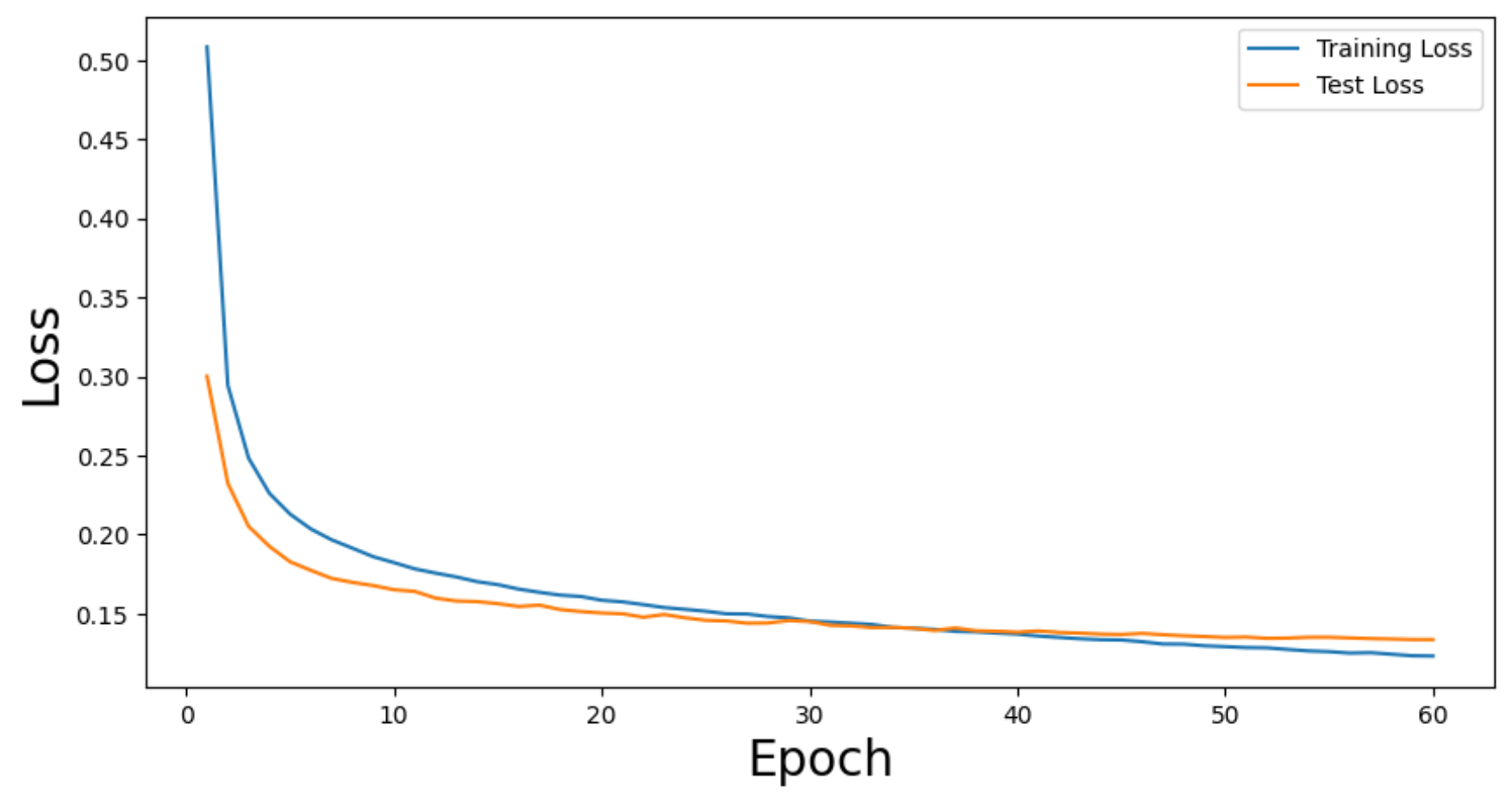}
    \caption{The loss values for training and test samples at each epoch. The model is trained for 60 epochs because the test loss value does not significantly decrease and starts to fluctuate.}
    \label{loss}
\end{figure}

The model is trained for a total of 60 epochs, as shown in Figure~\ref{loss}. After 60 epochs, the test loss values level off and begin to fluctuate, while the training loss continues to decrease. This indicates an increase in model complexity and a risk of overfitting. Therefore, we halt training at 60 epochs and complete the training process.

\begin{figure*}
    \centering
    \includegraphics[width=\linewidth]{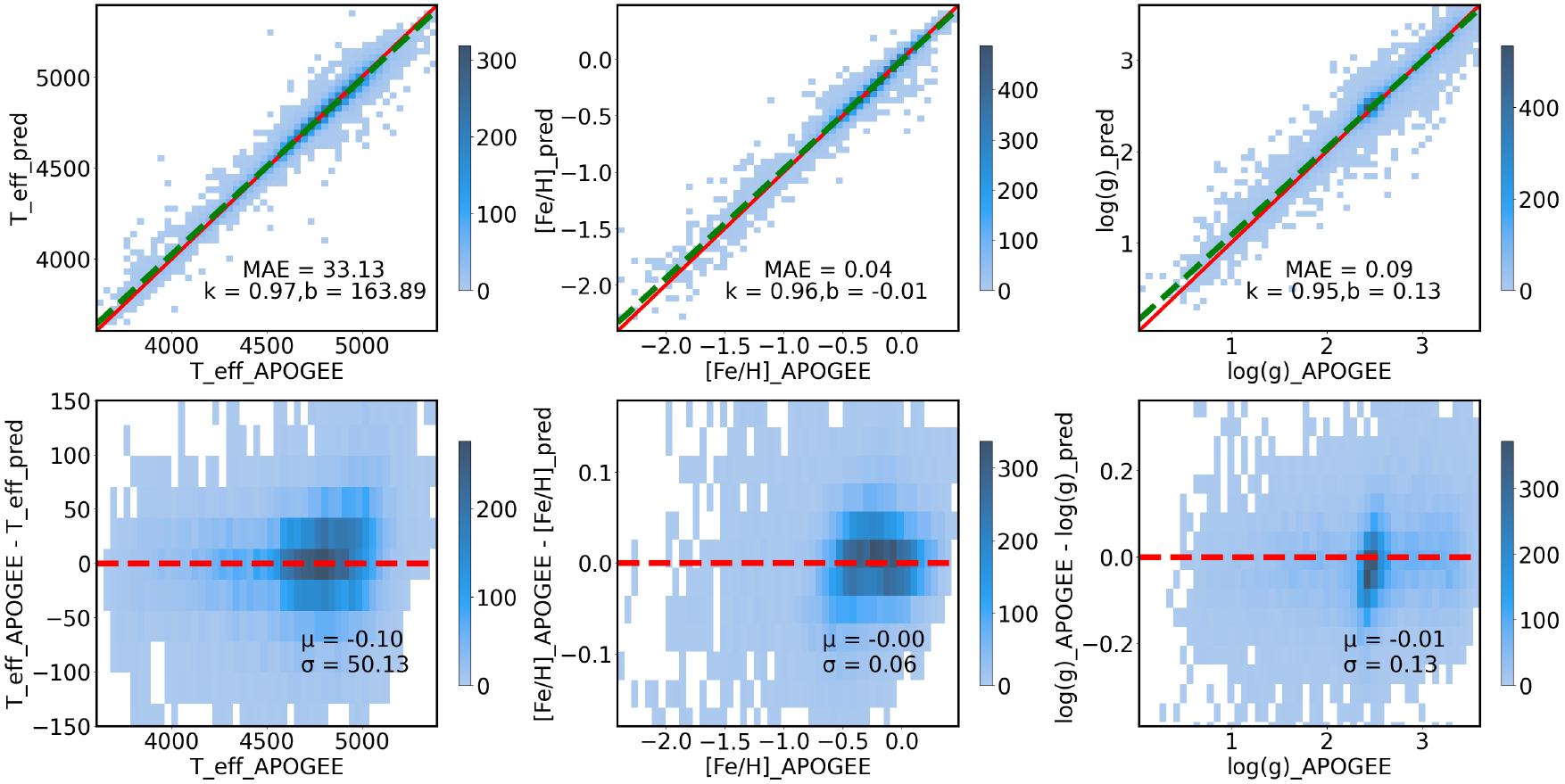}
    \caption{The upper row: the predicted labels vs. APOGEE labels in $T_{\text{eff}}$, [Fe/H] and log $g$. The red line represents a 1:1 ratio, while the green dashed line indicates the best-fit linear regression of the labels (k is the slope and b is the interception). The bottom row: the residuals along APOGEE labels within three standard deviations of the mean. The red dashed line indicates the residuals equal to 0. Note that Figures~\ref{stellarpara}, \ref{e1}, \ref{e2}, \ref{pre}, and \ref{com} all use linear density color maps. All the color bars in each panel of the stated figures represent the density.}
    \label{stellarpara}
\end{figure*}

\begin{figure*}
    \centering
    \includegraphics[width=1\linewidth]{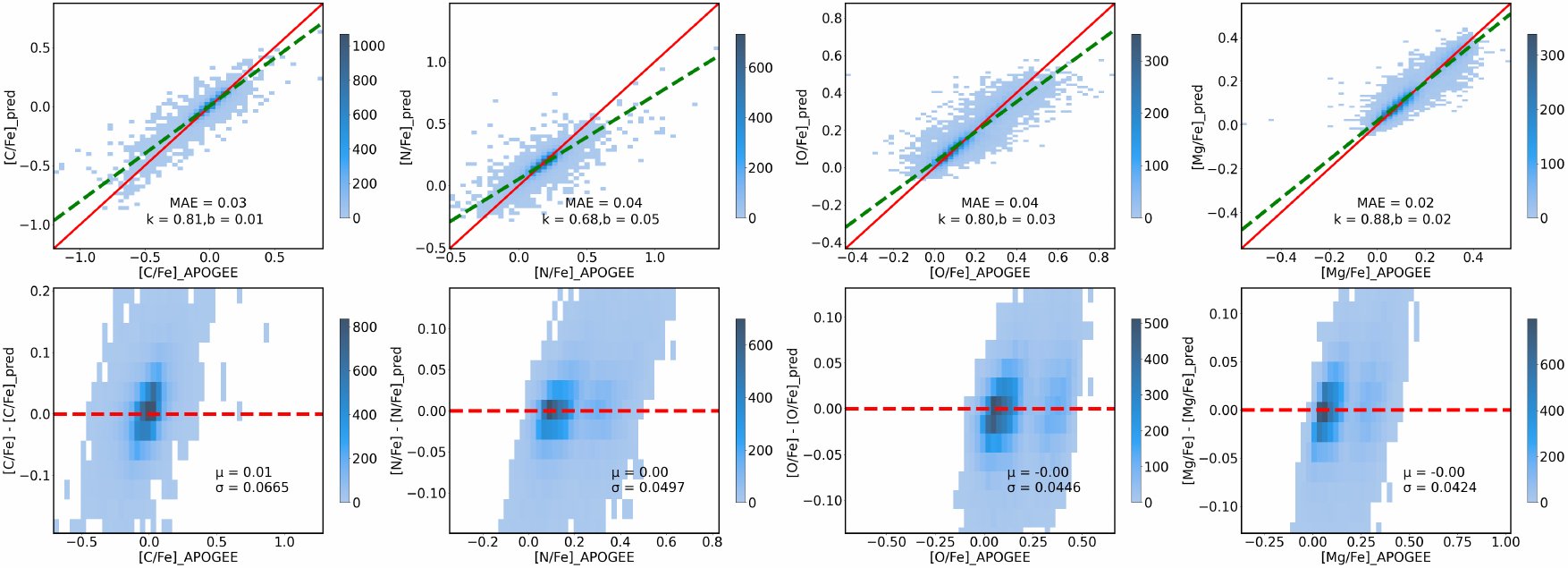}
    \caption{The upper row: the predicted labels vs. APOGEE labels in C, N, O and Mg elements. The red line represents a 1:1 ratio, while the green dashed line indicates the best-fit linear regression of the labels (k is the slope and b is the interception). The bottom row: the residuals along APOGEE labels within three standard deviations of the mean (with biases $\mu$ and scatters $\sigma$ given). The red dashed line indicates the residuals equal to 0.}
    \label{e1}
\end{figure*}

\begin{figure*}
    \centering
    \includegraphics[width=1\linewidth]{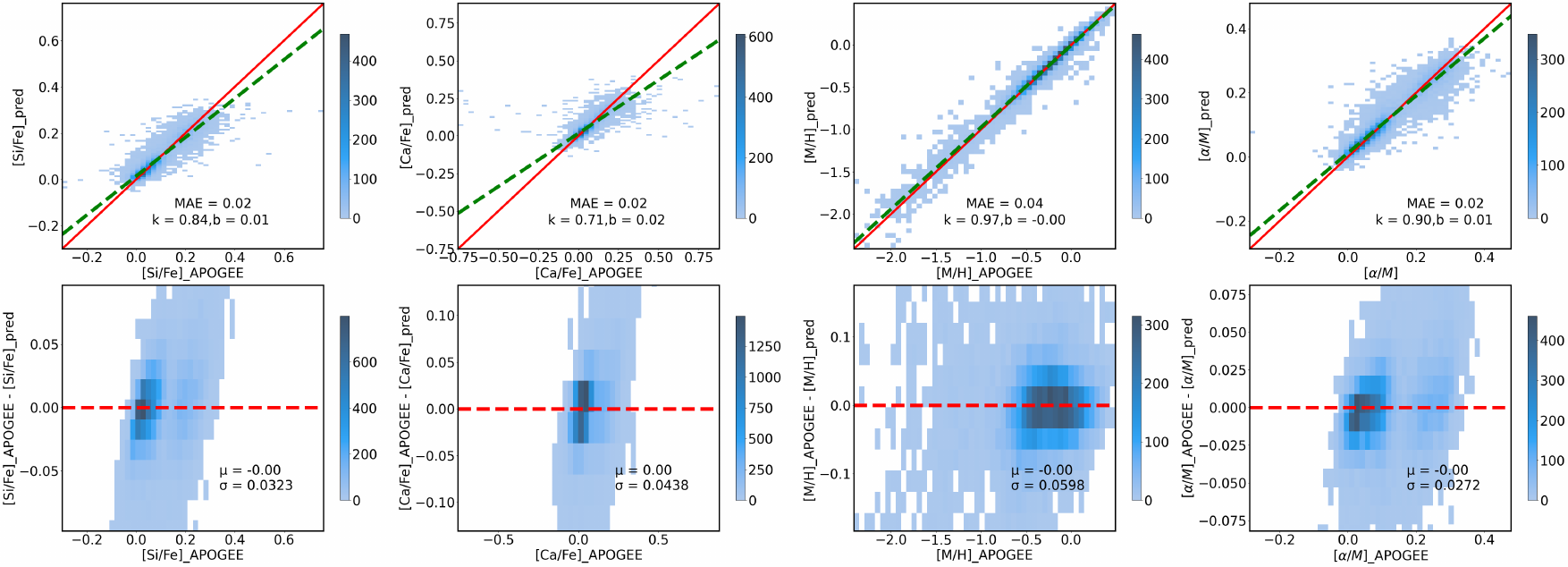}
    \caption{The upper row: the predicted labels vs. APOGEE labels in Si, Ca elements as well as overall metallicity and $\alpha$-abundance. The red line represents a 1:1 ratio, while the green dashed line indicates the best-fit linear regression of the labels (k is the slope and b is the interception). The bottom row: the residuals along APOGEE labels within three standard deviations of the mean (with biases $\mu$ and scatters $\sigma$ given). The red dashed line indicates the residuals equal to 0.}
    \label{e2}
\end{figure*}

\begin{figure*}
    \centering
    \includegraphics[width=\linewidth]{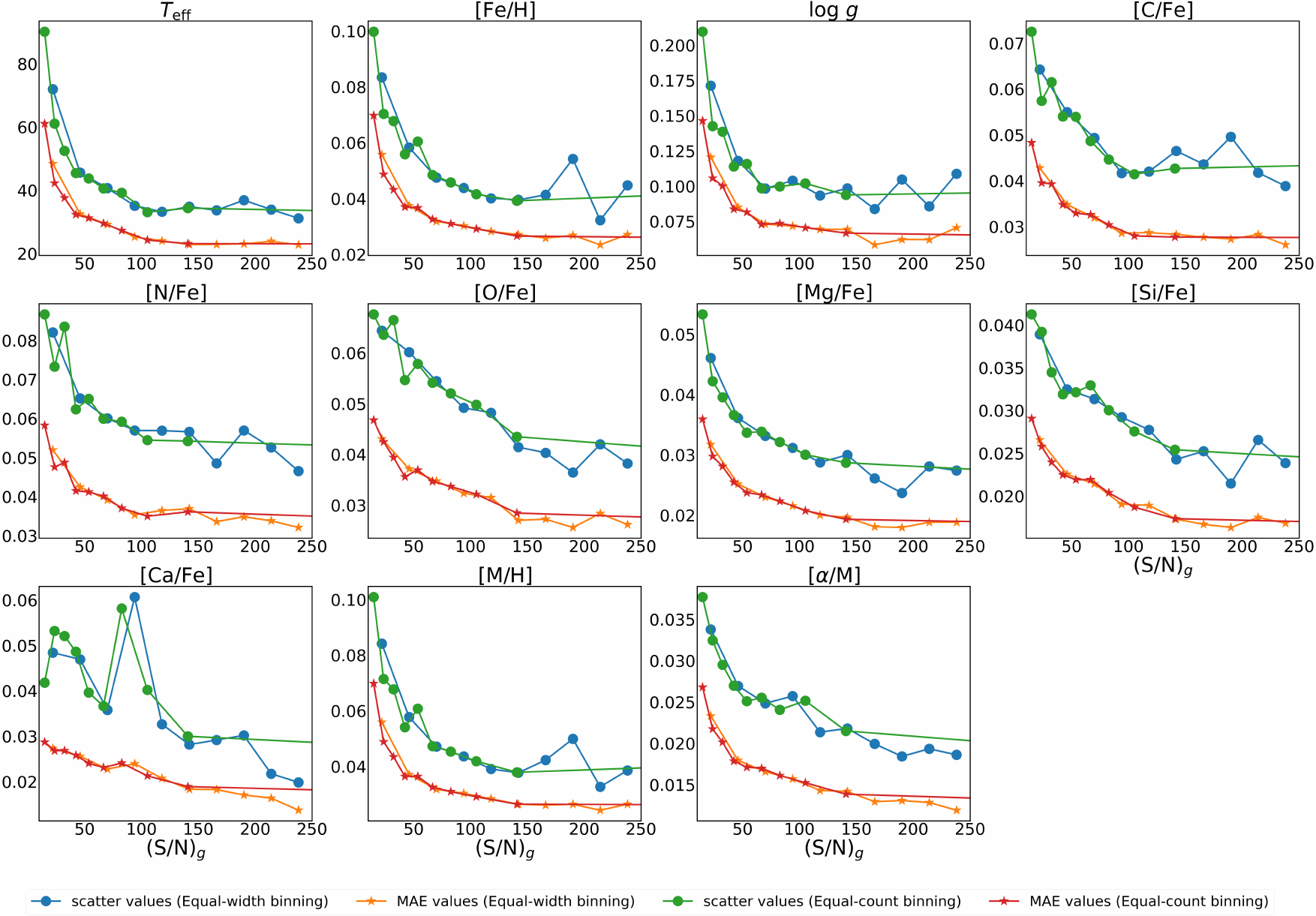}
    \caption{The dependence for scatter values (blue) and MAE values (orange) as a function of (S/N)$_{g}$ for each stellar label. The (S/N)$_{g}$ between 10 and 250 are evenly divided into 10 bins with scatter and MAE values calculated in each bin. Equal-count binning method is also applied for more scientific and better comparisons.} 
    \label{snr}
\end{figure*}

\subsection{Test Predictions}
We evaluate the test sample after the completion of training process and predict the stellar labels, which are $T_{\text{eff}}$, [Fe/H], log $g$, [C/Fe], [N/Fe], [O/Fe], [Mg/Fe], [Si/Fe], [Ca/Fe], overall metallicity [M/H] and $\alpha$-abundance [$\alpha$/M]. To better analyze the model performance, we combine MAE values with scatter (one standard deviation of the residuals; $\sigma$) and bias (mean values of the residuals; $\mu$) to explore the errors and data distributions. The results are shown in Figure~\ref{stellarpara}, Figure~\ref{e1} and Figure~\ref{e2}.

For $T_{\text{eff}}$, [Fe/H] and log $g$, the MAE values are 33.13 K, 0.04 dex and 0.09 dex respectively. The biases are $-0.10$ K for $T_{\text{eff}}$, 0 dex for [Fe/H] and $-$ 0.01 dex for log $g$. The corresponding scatters are 50.13 K, 0.06 dex and 0.13 dex respectively. These results are comparable with estimates for LAMOST DR8 giants from \citet{Li2022} using \texttt{astroNN}. The scatters in their work are $\sim$ 50.12 K for $T_{\text{eff}}$, $\sim$ 0.06 dex for [Fe/H] and $\sim$ 0.12 dex for log $g$. The log $g$ scatter is slightly lower in their work probably because they also include sub-giants in their samples and thus have more mapping relations between the spectra and log $g$ values. The MAE values are 0.03 dex for [C/Fe], 0.02 dex for [Mg/Fe] and 0.04 dex for both [N/Fe] and [O/Fe]. The biases for the elements N, O and Mg are both 0 dex, while the bias for C is 0.01 dex. The scatters are 0.07 dex for [C/Fe], 0.05 dex for [N/Fe], 0.04 dex for both [O/Fe] and [Mg/Fe]. Both [Si/Fe] and [Ca/Fe] have MAE values of 0.02 dex and 0 dex biases, with scatters of 0.03 dex and 0.04 dex respectively. For overall metallicity and $\alpha$-abundance, the MAE values are 0.04 dex and 0.02 dex with biases of both 0 dex and scatters of $0.06$ dex and $0.03$ dex respectively.

The functional relationships between the scatter values and MAE values with respect to (S/N)$_{g}$ are exhibited in Figure~\ref{snr}. Both scatter and MAE values generally show a downward trend with increasing (S/N)$_{g}$. However, fluctuations can be observed at high (S/N)$_{g}$ values, probably because there are fewer points with high (S/N)$_{g}$. Another interesting fact is that there is a prominent peak in the [Ca/Fe] scatter value in the interval (82, 106) of (S/N)$_g$, indicating that there are more predicted outliers and that the scatter value is more sensitive to these outliers. Similar situation also happened when \citet{Li2023} analyzed the uncertainty of [Cr/H].

\subsection{Comparisons with Other Works}
Despite similar precisions in $T_{\text{eff}}$, [Fe/H] and log $g$ with the work from \citet{Li2022} as mentioned above, our scatters are also comparable to theirs, although for most elements, the scatters are slightly higher (with an order of one-thousandth). Their lower scatters are attributed to their larger training sample size (more than 61,000 stars), which provides a greater number of mapping relationships. However, during their sample selection process, they did not filter the samples based on elemental errors or quality. As a result, even though their scatters are lower, the true values used in their training may still be problematic. \citet{zhang2019} uses StarNet to estimate the stellar parameters and abundances of giants from LAMOST DR5, reporting slightly smaller scatters in stellar parameters compared to ours. Their scatter for [M/H] is only 0.05 dex, which is one-third of our [M/H] scatter value. There are two possible reasons for this discrepancy: firstly, they impose stricter quality criteria when selecting APOGEE stars (i.e., requiring \texttt{STARFLAG} = 0 and \texttt{ASPCAPFLAG} = 0); secondly, there may be methodological errors in mapping the spectra to [M/H] values. \citet{Ho2017} employed \texttt{the Cannon} method to estimate stellar parameters and abundances for giants from LAMOST DR2, achieving scatters of 70 K in $T_{\text{eff}}$, 0.1 dex in both log $g$ and [Fe/H], 0.04 dex in [$\alpha$/M] with (S/N)$_{g} >$ 50. Our results are comparable to their uncertainties, considering that our test samples also include stars with (S/N)$_g$ lower than 50. We also notice that \citet{Wang2022} only used the wavelength ranges of 3900–5800 Å and 8450–8950 Å as input spectra for predicting LAMOST labels, due to the potential for significant background contamination from sky emission lines and telluric bands in the intermediate wavelength range. However, our NN method demonstrates that incorporating the information from 5800–8450 Å does not adversely affect the mapping relations between the spectra and elemental abundances. In fact, our derived [C/Fe] and [N/Fe] ratios exhibit lower scatters than their estimations. These results suggest that our model is robust and can reliably predict stellar labels.

\subsection{The Validation of Uncertainties}
\citet{Gal2015} developed a new theoretical framework casting dropout training in deep NN as approximate Bayesian inference in deep Gaussian processes, which could reduce the computational cost that occurs in Bayesian networks when modeling the uncertainties. This idea is applied by both \citet{Leung2019} and \citet{Li2023}. Activating dropout layers during inference introduces stochasticity by randomly deactivating neurons in each forward pass. Repeated predictions for the same input generate different sub-networks, producing a distribution of outputs. The variance across these predictions quantifies epistemic uncertainty (model uncertainty due to parameter estimation). As a result, we used the standard variance of each predicted label to represent the uncertainties in a trained model with dropout layers activated. We estimated the parameters 10 times to obtain more consistent and robust uncertainties, which is twice the number used in \citet{Li2023}.  The choice of this method here offers a simple, low-cost way to estimate model uncertainty through simple stochastic forward passes, without requiring architectural changes or retraining. However, it primarily captures epistemic uncertainty while underestimating input noise effects and cannot account for instrumental uncertainties. A recent study shows that NN models often struggle with input noise \citep{V2025}, so this method is straightforward and avoids large uncertainty space that is hard to quantify.

\begin{figure*}
    \centering
    \includegraphics[width=1\linewidth]{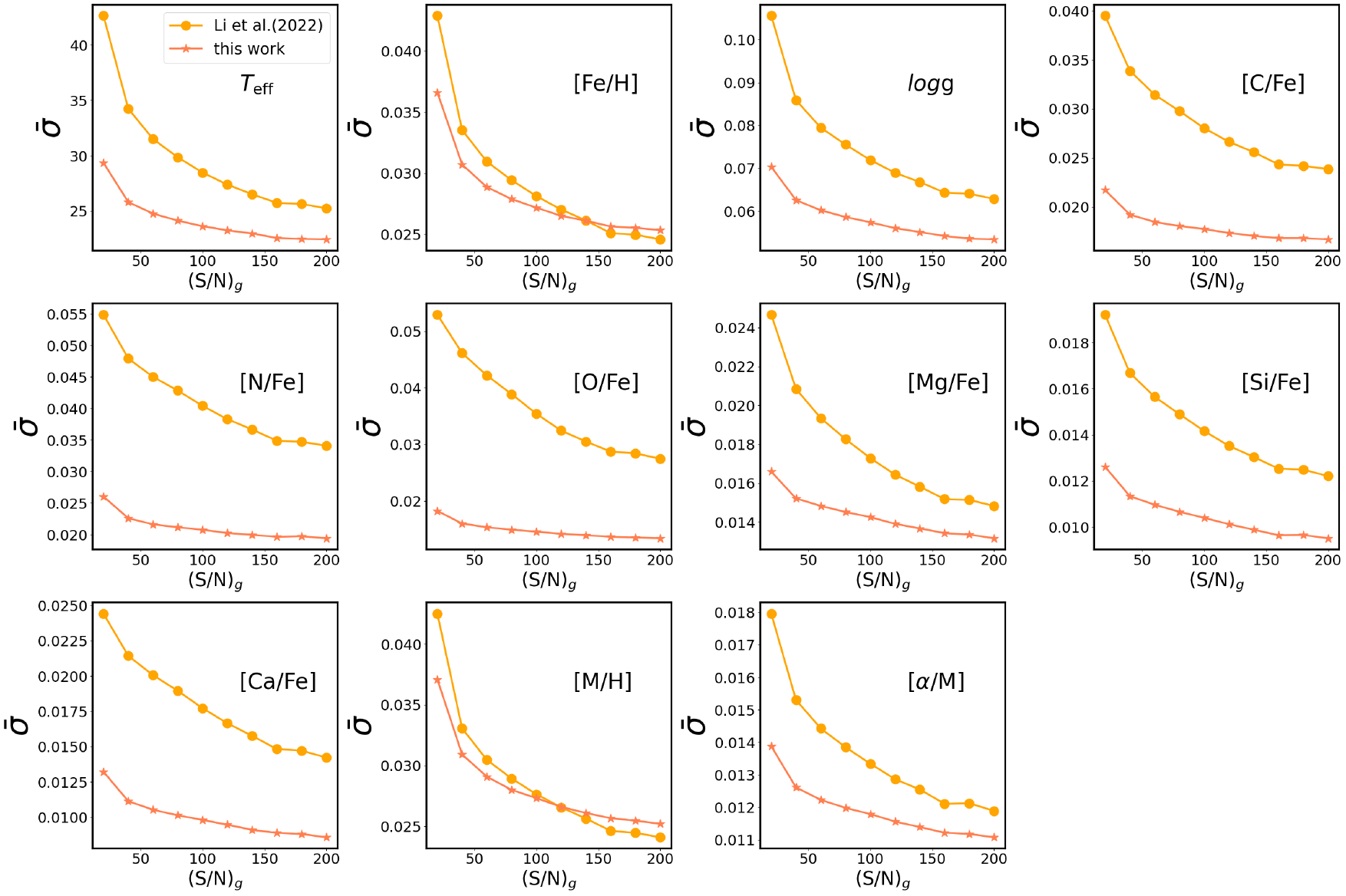}
    \caption{The comparison of the mean prediction uncertainties for common stars between our work and \citet{Li2022} reveals that our model exhibits significantly lower uncertainties in most cases. However, the two models show comparable uncertainties in (overall) metallicity. Notably, our model performs slightly better when (S/N)$_g \lesssim 100$. It should be noted that the uncertainties from both studies have been rounded to two decimal places.}
    \label{uncertainty}
\end{figure*}

Figure~\ref{uncertainty} compares the uncertainties of our model and \citet{Li2022} (respectively) across 10 (S/N)$_g$ bins, ranging from 10 to 210. Prediction uncertainties show a downward trend with increasing (S/N)$_g$. In most estimated parameters and elemental abundances, our model has lower uncertainties at the low (S/N)$_g$ end. The uncertainties of almost all elemental abundances (except for N) in our model are $\lesssim$ 0.02 dex, and the uncertainties of the overall $\alpha$-abundances are also lower than those in \citet{Li2022}. Besides, the two models show compatible uncertainties in metallicity and overall metallicity, and our model performs slightly better when (S/N)$_g \lesssim 100$, while theirs performs slightly better when (S/N)$_g \gtrsim 100$.

\begin{figure*}
    \centering
    \includegraphics[width=1\linewidth]{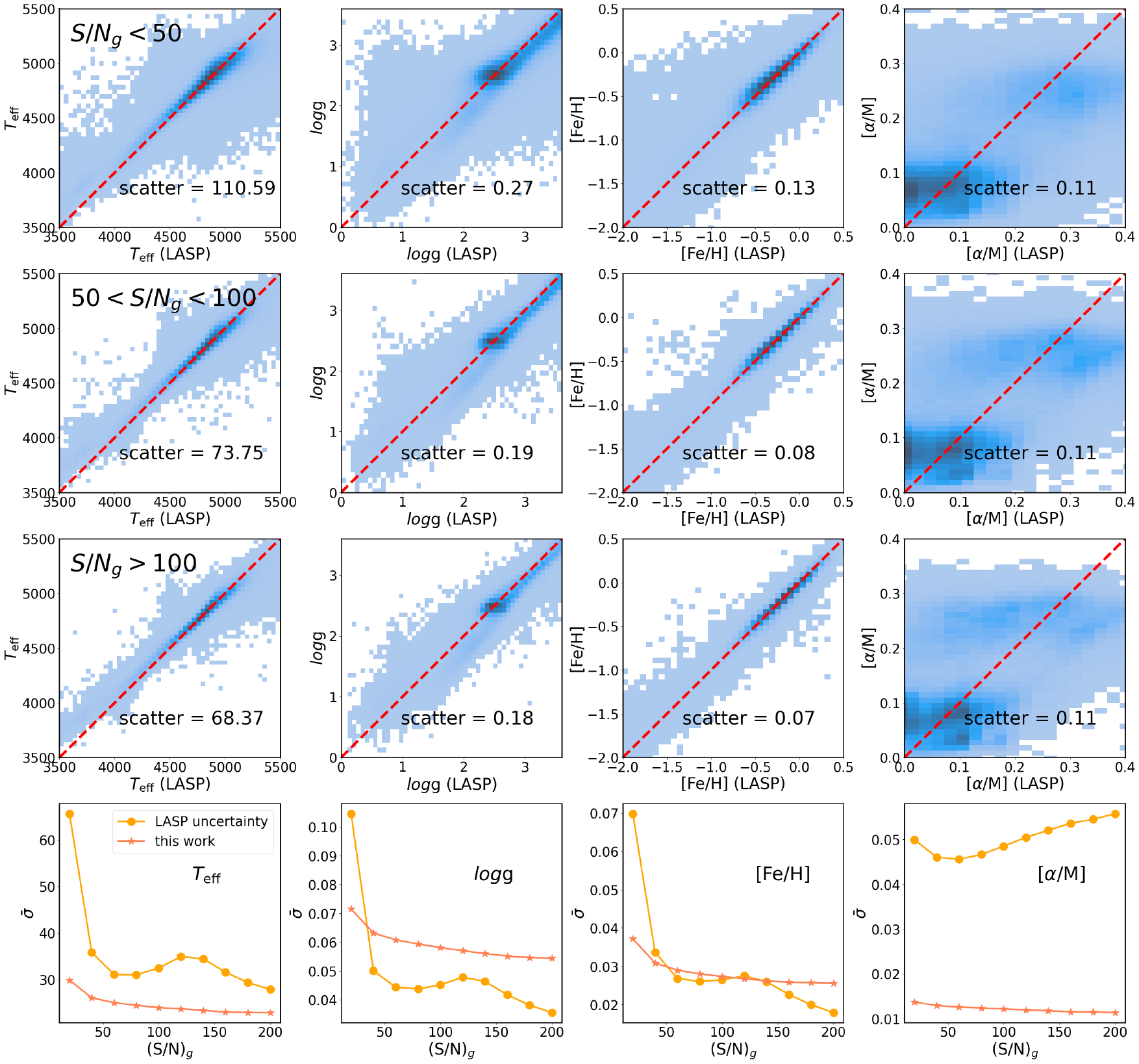}
    \caption{The comparison between LASP estimates and predictions in our work. Top three rows: comparisons of $T_{\text{eff}}$, $log$g, [Fe/H] and [$\alpha$/M] as a function of (S/N)$_g$. These rows do not include a color bar, as they are meant only for comparative purposes. Bottom row: comparison of mean uncertainties with increasing (S/N)$_g$. }
    \label{lasp}
\end{figure*}

To further validate the uncertainties, we compared the prediction results with the values (i.e., $T_{\text{eff}}$, $log$g, [Fe/H] and [$\alpha$/M]) and corresponding uncertainties that are available from LASP (see Figure~\ref{lasp}). As shown in the first three rows of Figure~\ref{lasp}, the scatter values of all parameters, except for [$\alpha$/M], decrease with increasing (S/N)$_g$. Among them, the scatter value of $T_{\text{eff}}$ exhibits the most significant decline. The scatter value of [$\alpha$/M] remains unchanged with increasing (S/N)$_g$, which is likely attributable to the high LASP uncertainties of [$\alpha$/M] (see bottom row). These uncertainties even exhibit an upward trend at the high (S/N)$_g$ end, resulting in a large and sparse distribution in the LASP [$\alpha$/M] values. The bottom row of Figure~\ref{lasp} presents a comparison of between prediction uncertainties and LASP uncertainties as a function of (S/N)$_g$, and it is consistent with Figure~\ref{uncertainty}. For $T_{\text{eff}}$ and [$\alpha$/M], the uncertainties in this work are significantly lower than those of LASP, particularly for [$\alpha$/M]. However, the mean uncertainties of our work for surface gravity are only lower when (S/N)$_g \lesssim 50$. For metallicity, the uncertainties in our work are comparable to LASP uncertainty estimates in the range $50 \lesssim \text{(S/N)}_g \lesssim 150$. Our model performs notably better at the low (S/N)$_g$ end, while LASP provides better estimates towards the high (S/N)$_g$ end.

Finally, users can select trustworthy elemental abundances by applying constraints on the uncertainties. When selecting stars based on $T_{\text{eff}}$, special care must be taken for cool ($<4000$ K) and hot ($>5000$ K) stars, as these exhibit relatively large $T_{\text{eff}}$ uncertainties, reaching up to $\sim50$ K. For stars with predicted $T_{\text{eff}}$ between 4000 K and 5000 K, the uncertainties are all below 30 K. Besides, stars with lower surface gravity (i.e., $\log g \lesssim 1.4$ dex) and stars that are more metal-poor also exhibit larger corresponding uncertainties (for more details, see Subsection~\ref{vmp}). As a result, users can utilize LASP estimates of surface gravity (if available) when constraining the desired samples. It is also important to emphasize that the tested elemental abundances fall within the coverage of the training sets for nearly all stars in the sample. However, one star in the test set, with the observation ID (\texttt{obsid}) 92705131, exhibits [Mg/Fe] values that lie outside the range of the training set. This leads to a significant extrapolation issue, rendering the predicted labels for this star unreliable, so it is strongly recommended that users rely on the ASPCAP estimates for this particular star instead.

It is also important to emphasize that the uncertainties discussed here do not directly correlate with the offsets between the predicted labels and the APOGEE labels. This is because the uncertainties of NN models pertain to the stability of the predictions rather than the accuracy of the offsets. The APOGEE labels with small values tend to be overestimated, while those with high values are often underestimated (e.g., Figure~\ref{pre} and Figure~\ref{com}). This issue was also observed when using LAMOST medium-resolution spectra \citep{Wang2023}. Although the predicted labels at the ends have small uncertainties, the offsets could be large. The possible reason for this is that lower abundances result in less prominent spectral features, making them difficult for the NN to detect, while features with higher abundances may be flattened during preprocessing. Therefore, we predicted all the common stars and calculated the standard deviations of the offsets as $\sigma$. For predicted labels with offsets within ($-\sigma$,$\sigma$), the corresponding ranges are: 3645 K $<T_{\text{eff}}<$ 5365 K, $-2.31$ dex $<$ [Fe/H] $<0.44$ dex,  0.38 dex$<log$g$<3.62$ dex, $-0.75$ dex $<$ [C/Fe] $<0.47$ dex, $-0.24$ dex $<$ [C/Fe] $<0.82$ dex, $-0.07$ dex $<$ [O/Fe] $<0.53$ dex, $-0.05$ dex $<$ [Mg/Fe] $<0.45$ dex, $-0.04$ dex $<$ [Si/Fe] $<0.32$ dex, $-0.10$ dex $<$ [Ca/Fe] $<0.39$ dex, $-2.31$ dex $<$ [M/H] $<0.46$ dex, $-0.04$ dex $<$ [$\alpha$/M] $<0.34$ dex. Within these ranges, at least $\sim79\%$ stars have offsets $<1\sigma$, and $\sim98\%$ of stars have offsets $<3\sigma$, which can serve as a reliable reference.

The underestimation of uncertainties remains largely unavoidable in data-driven methods. \citet{Huang2020} quantified the age uncertainties of red clump stars by combining both random errors (the uncertainties in our work) and systematic method errors (the scatter values in our work) in quadrature: $\sigma_{\mathrm{total}} = \sqrt{\sigma^2_{\mathrm{random}} + \sigma^2_{\mathrm{method}}}$. Although this approach can partially mitigate the underestimation of uncertainties,  $\sigma_{\mathrm{method}}$ (i.e., scatter values) remain relatively small compared to the real offsets (i.e., the reduction in underestimation is insignificant). For example, there are 166 stars with (S/N)$_g=10$ and the median uncertainties for [Mg/Fe] are $\sim0.02$ dex and $\sim0.05$ dex, respectively, based on the two definitions of uncertainty. We therefore still recommend using the elemental abundance ranges provided above, despite having computed uncertainties through this formalism.

\section{Predictions For LAMOST DR10 Giants}\label{predictions}
\subsection{Distributions of Predicted Stellar Labels}\label{vmp}
\begin{figure*}
    \centering
    \includegraphics[width=1\linewidth]{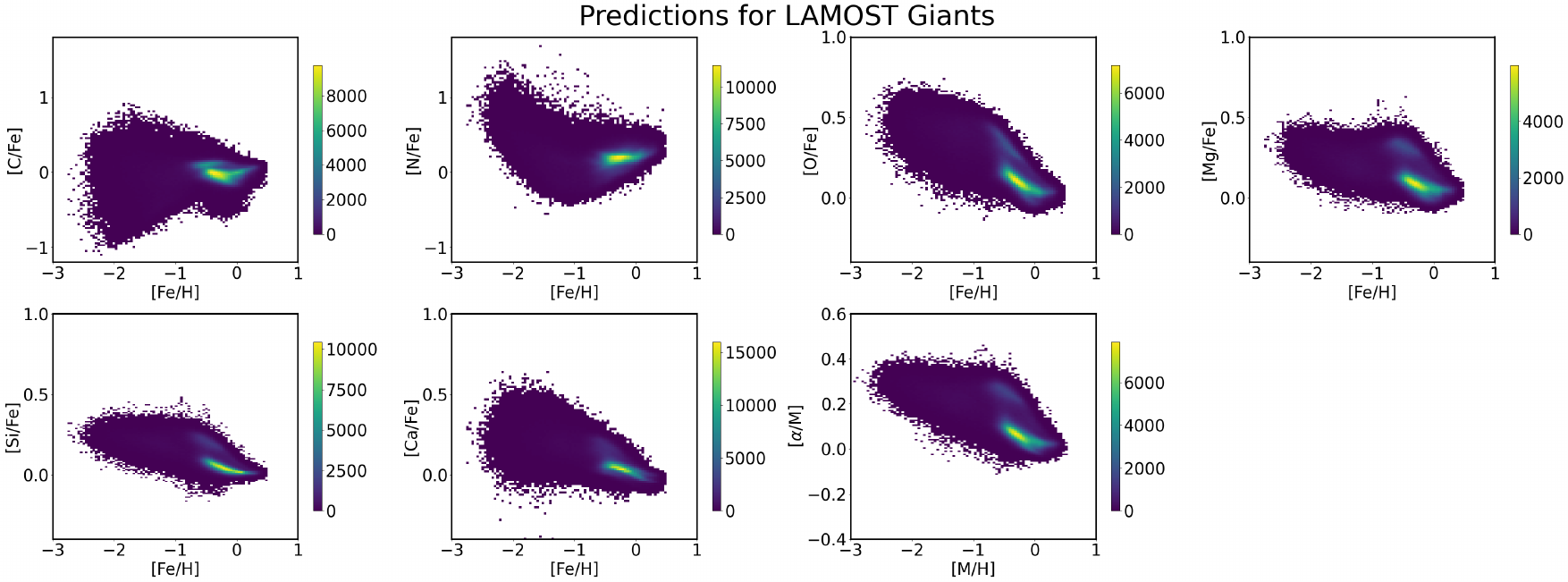}
    \caption{Predicted elemental labels for all LAMOST DR10 giants and corresponding elemental distributions. The bimodal structure of Galactic disks is clearly observed, with the high-$\alpha$ clump corresponding to the thick disk and the low-$\alpha$ clump representing the thin disk.}
    \label{pre}
\end{figure*}

\begin{figure*}
    \centering
    \includegraphics[width=1\linewidth]{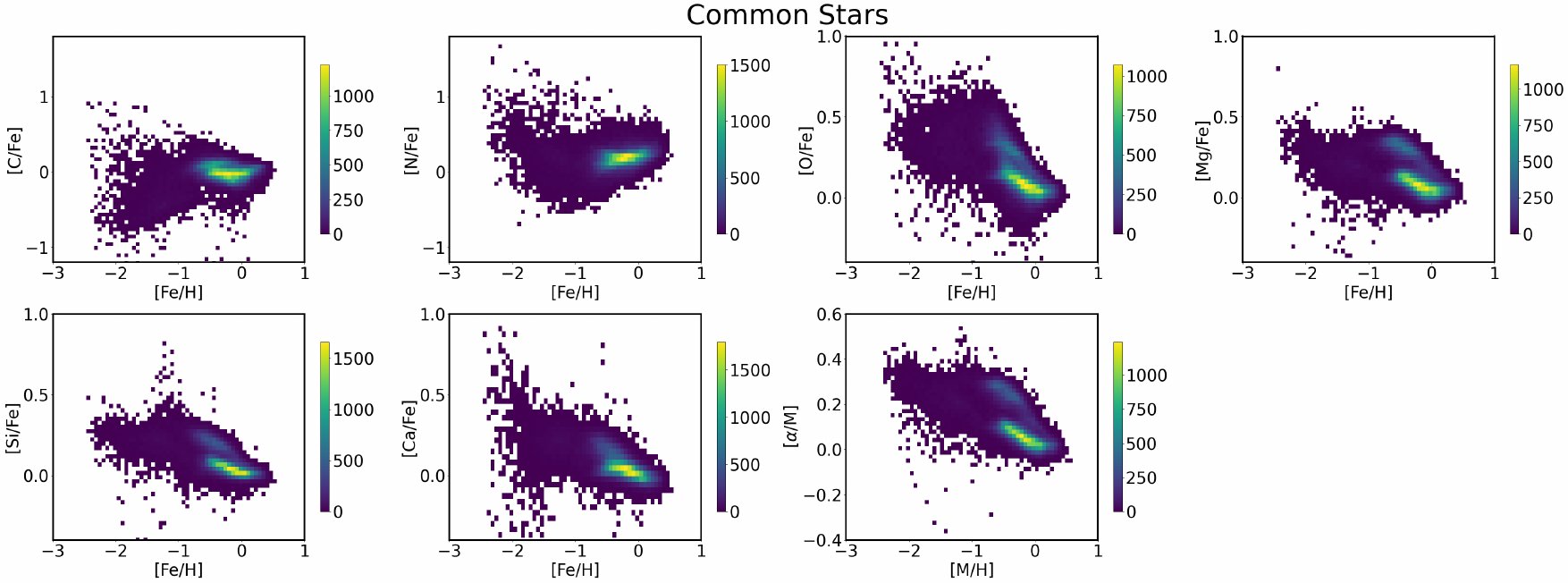}
    \caption{Elemental distributions of common stars for comparison with the predicted distributions for LAMOST giants. Note that abundances here are estimated from APOGEE.}
    \label{com}
\end{figure*}

We normalized and standardized the spectra of 1,100,858 giants from LAMOST DR10 as input, predicting the stellar labels for each star. The distributions of the predicted labels are shown in Figure~\ref{pre}, along with the distributions of common stars for better comparison in Figure~\ref{com}. The bimodal distributions are clearly observed in all [X/Fe]-[Fe/H] planes (where X denotes C, N, O, Mg, Si, Ca) as well as [$\alpha$/M]-[M/H] plane. The two distinct clumps in these planes are associated with the high-$\alpha$ thick disk and low-$\alpha$ thin disk components \citep{G1983,F1998,Bensby2014}.

The predicted stellar labels have following ranges: $-$2.76 dex $<$ [Fe/H] $<$ 0.50 dex, $-$2.96 dex $<$ [M/H] $<$ 0.54 dex, $-$1.11 dex $<$ [C/Fe] $<$ 0.93 dex, $-$0.57 dex $<$ [N/Fe] $<$ 1.69 dex, $-$0.14 dex $<$ [O/Fe] $<$ 0.75 dex, $-$0.10 dex $<$ [Mg/Fe] $<$ 0.63 dex, $-$0.17 dex $<$ [Si/Fe] $<$ 0.49 dex, $-$0.41 dex $<$ [Ca/Fe] $<$ 0.64 dex and $-$0.11 dex $<$ [$\alpha$/M] $<$ 0.46 dex. The metal-poor stars with [Fe/H] $< -$1.0 account for approximately 6\% of the entire sample (65,120 stars), while the very metal-poor stars with [Fe/H] $< -$2.0 constitute only $\sim$ 0.3\% of the total sample (3,686 stars). \citet{Xie2021} shows that metal-poor stars have larger scatters than normal stars. To examine this, we calculate the elemental scatters for the metal-poor (MP) stars ($-$2.0 dex $<$ [Fe/H] $< -$1.0 dex) and very metal-poor (VMP) stars ([Fe/H] $< -$ 2 dex). For the MP stars, the scatters are 0.14 dex for [Fe/H], 0.15 dex for C, 0.20 dex for N, 0.11 dex for O, 0.09 dex for Mg, 0.07 dex for Si, 0.15 dex for Ca and 0.06 dex for [$\alpha$/M]. And for the VMP stars, the scatters are 0.17 dex for [Fe/H], 0.29 dex for C, 0.40 dex for N, 0.32 dex for O, 0.08 dex for Mg, 0.04 dex for Si, 0.48 dex for Ca and 0.07 dex for [$\alpha$/M]. Despite the large elemental scatters among the MP and VMP stars, the scatters for Mg, Si and overall $\alpha$-abundance still remain below 0.10 dex. Thus, the three abundances can provide reliable information for studying the chemical evolution of MP and VMP stars. Additionally, the abundance data for these metal-poor stars will offer more insights into the accretion history when using these giants as tracers to study the Galactic halo.

\begin{figure*}
    \centering
    \includegraphics[width=\linewidth]{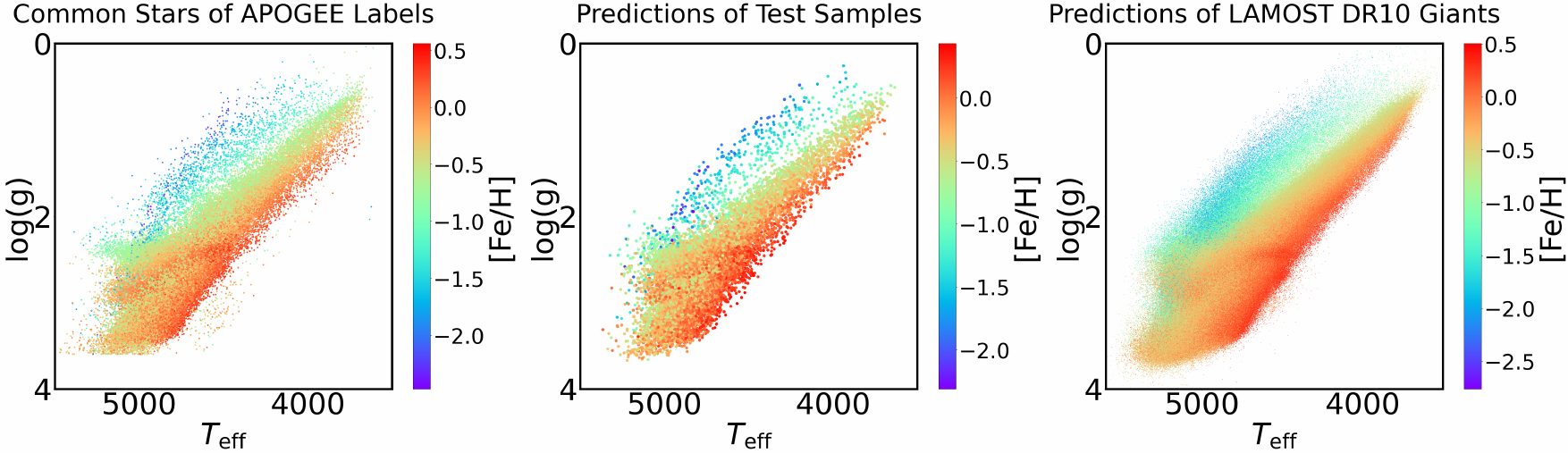}
    \caption{The Kiel diagrams for common stars of APOGEE label, test samples of predicted labels and LAMOST DR10 giants of predicted labels color-coded by metallicity. The predicted labels for LAMOST giants well recovered the main features in Kiel diagram. }
    \label{HR}
\end{figure*}

Figure~\ref{HR} shows the Kiel diagrams of the common stars, test samples and LAMOST giants color-coded by metallicity. The predicted labels for LAMOST giants successfully recover the position and slope of the red clump (RC) stars, as well as the smooth metallicity transition along the giant branch. The relatively metal-poor RC stars tend to have lower log $g$ values \citep{Zhao2001}. The predicted $T_{\text{eff}}$ ranges from 3468 K to 5622 K and log $g$ ranges from $-$0.29 dex to 4.13 dex, which generally coincides with the estimated values from LSAP. These consistent predicted labels further confirm the precision and ability of our CNN model.

At last, we provide a catalog of estimated stellar parameters and elemental abundances for $\sim$ 1.1 million giants from LAMOST DR10. The descriptions of the catalog are shown in Table~\ref{catalog}. In this catalog, we also include the geometric distances \texttt{r\_med\_geo} estimated by \citet{B2021} for more convenient calculation of stellar positions, velocities, actions, and other related parameters. For quantities missing values, we use NaN values for supplementation.

\begin{table*}
\caption{The description for each column in the provided catalog}
    \centering
    \begin{tabular}{|c|p{0.7\textwidth}|}  
      \hline
      Column Name & Descriptions \\
      \hline
      obsid & unique spectra ID from LAMOST \\
      \hline
      source\_id& source\_id from Gaia DR3 catalog, the value would be -9999 if the star is not available in Gaia DR3 catalog\\
      \hline
      ra& right ascension from LAMOST\\
      \hline 
      dec& declination from LAMOST\\
      \hline
      T\_eff& effective temperature estimated by CNN in this work\\
      \hline
      [Fe/H]& metallicity estimated by CNN in this work\\
      \hline
      logg& surface gravity estimated by CNN in this work\\
      \hline
      [C/Fe]& carbon abundance estimated by CNN in this work\\
      \hline
      [N/Fe]& nitrogen abundance estimated by CNN in this work\\
      \hline
      [O/Fe]& oxygen abundance estimated by CNN in this work\\
      \hline
      [Mg/Fe]& magnesium abundance estimated by CNN in this work\\
      \hline
      [Si/Fe]& silicon abundance estimated by CNN in this work\\
      \hline
      [Ca/Fe]&  calcium abundance estimated by CNN in this work\\
      \hline
      [M/H]& overall metallicity estimated by CNN in this work\\
      \hline
       [alpha/M]& overall $\alpha$-abundance estimated by CNN in this work\\
      \hline
      teff\_lasp& effective temperature estimated by LASP\\
      \hline
      logg\_lasp& surface gravity estimated by LASP \\
      \hline
      r\_med\_geo& the median of the geometric distance posterior estimated by \citet{B2021}\\
      \hline
     X\_sigma & X represents teff, feh, logg, cfe, nfe, ofe,mgfe, sife, cafe, mh and am, corresponding to the uncertainties of:  effective temperature, metallicity, surface gravity, C, N, O, Mg, Si, Ca abundances,  overall metallicity, and overall $\alpha$-abundances. \\
     \hline
     X\_sigma\_h & The same denotations as the above row, but with uncertainties calculated using the method in \citet{Huang2020} \\
      \hline
    in\_APOGEE & 1 if the star is in APOGEE; 0 if the star is not in APOGEE. \\
      \hline
    \end{tabular}
    \label{catalog}
\end{table*}

\subsection{The Elemental Distributions of Discovered Substructures}

The substructures in the Galactic stellar halo typically refer to the clumps in kinematic and chemical spaces. Most of these substructures are accreted from dwarf galaxies, while others may have in-situ origins but exhibit different kinematic or chemical patterns compared to the in-situ stellar halo. One example is the Splash \citep{belokurov2020}, a metal-rich and highly eccentric substructure containing proto-disk stars that were heated by the last major merger event. The last major merger event is commonly referred to as the Gaia-Sausage-Enceladus (GSE) merger \citep{Berokurov2018,Helmi2018}, which occurred $\sim8-10$ Gyr ago \citep{Berokurov2018,Gallart2019}. The member stars of the GSE exhibit a distinct sausage-shaped or blob-shaped distribution in $V_r-V_{\phi}$ space, with the majority of members having eccentricities exceeding 0.7.  Very retrograde and high-energy substructures such as Sequoia are distinct from GSE members with higher [Al/Fe] values, whose progenitor is estimated to have a total mass of $\sim10^8{M_\odot}$ \citep{Myeong2019}. However, simulations also show that a GSE-like merger could generate Sequoia-like substructure in $E-L_z$ space (i.e., see Amarante et al.\citeyear{Amarante2022}).

In the previous work, \citet{Liu2024} have identified three possibly new substructures in integral-of-motions space using HDBSCAN clustering algorithm, which includes orbital energy ($E$), angular momentum z-component ($L_z$) and angular momentum in-plane component ($L_{\perp}=\sqrt{L_x^2+L_y^2}$). The substructures are Prograde Substructure 1 (PG1), Prograde Substructure 2 (PG2) and the Low Energy Group (LEG). PG1 is considered either as a low-eccentricity, metal-rich component of GSE or as part of the metal-poor disk, while PG2 is regarded as its relatively high-eccentricity component or a mixture of Aleph and the disk, due to its kinematic similarities with Aleph \citep{Naidu2020}. In contrast, LEG exhibits very low orbital energy, and its member stars have extremely high eccentricities approaching $\sim$1. LEG also comprises multiple stellar populations, as its member stars are predominantly located in the inner region of the Galaxy (i.e., the Bulge region). Additionally, it includes the Heracles substructure \citep{Horta2021}, which was identified through a Gaussian Mixture Model (GMM) analysis of the metallicity distribution function (MDF). Further analysis is hindered because these substructures were identified using \textit{Gaia} DR3, and they lack detailed elemental abundance analysis. Nevertheless, a significant advantage of using \textit{Gaia} DR3 is its potential to uncover additional substructures, given the vast amount of high-quality data.

\begin{figure*}
    \centering
    \includegraphics[width=\linewidth]{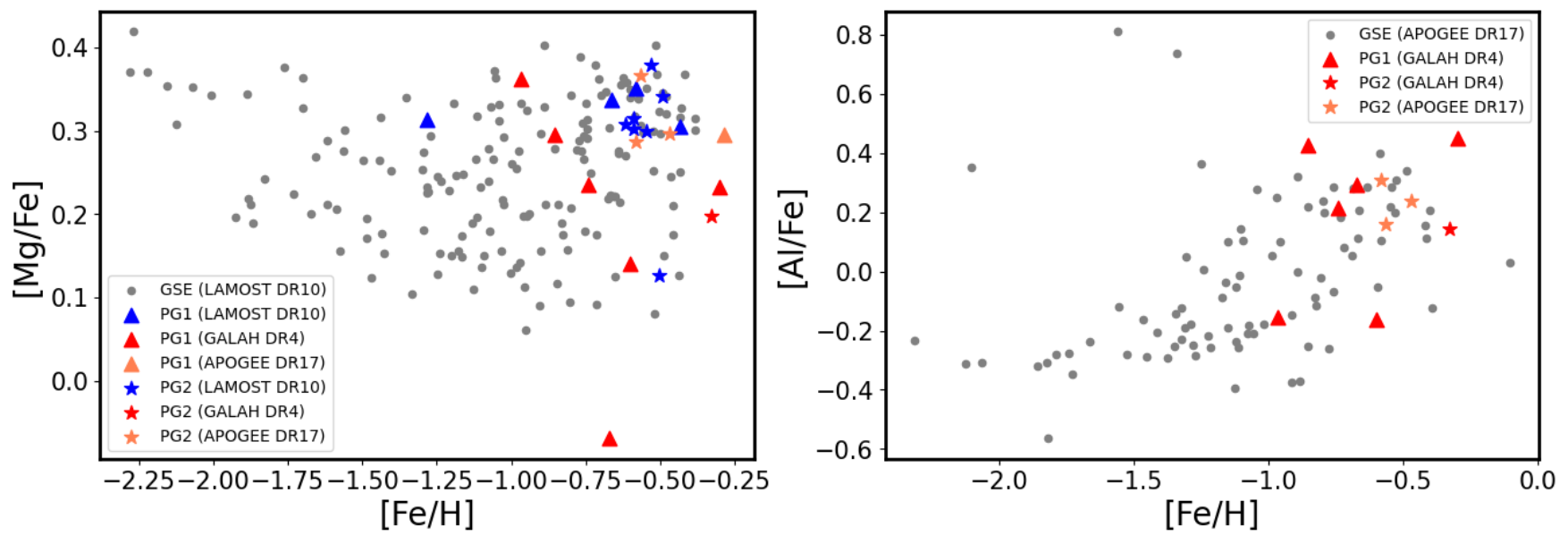}
    \caption{The selected GSE, PG1 and PG2 common stars are from \citet{Liu2024}. Left panel: the distributions in [Mg/Fe]-[Fe/H] plane. Gray dots represent GSE member stars with elemental abundance estimates from LAMOST DR10. Triangles denote PG1, with blue, red, and coral colors indicating estimates from LAMOST DR10, GALAH DR4, and APOGEE DR17, respectively. The same color scheme and a symbol of star are used for PG2 as for PG1. Right panel: the gray dots represent GSE members with elemental abundance estimates from APOGEE DR17. Only GALAH DR4 provides [Al/Fe] values for PG1, represented by red triangles. For PG2, red and coral triangles correspond to estimates from GALAH DR4 and APOGEE DR17, respectively. Note that some metal-rich ([Fe/H]$>-1$) GSE stars have high [Mg/Fe] values ([Mg/Fe]$>0.2$) also exhibit positive [Al/Fe] values, suggesting that these stars could be the contamination from Splash.}
    \label{mg}
\end{figure*}

\citet{Dodd2024} independently identified LEG substructure (referred to as ``Low Energy (LE)" in their study) using data from \textit{Gaia} DR3 and APOGEE DR17. In the [Mn/Mg]-[Al/Fe] plane, they discovered that LEG consists of a mixture of both accreted and in-situ populations, with the accreted component confirming the presence of the Heracles substructure within LEG. According to their study, the in-situ population of LEG is intrinsically old, with half of its stars having formed about 12.9 Gyr ago. The LEG identified in both studies exhibits low $L_z$ values ($<500$ kpc km/s), indicating high eccentricity. Additionally, LEG in both studies exhibits similar metallicity distribution functions, peaking at [Fe/H]$\sim-0.6$. The differences in $L_{\perp}$ are likely due to the discrepancy of adopted potential model and clustering algorithm. Given that both studies detected LEG in the same integral space of motion (i.e., $E, L_Z$ and $L_{\perp}$), we think that the observed LEG is fundamentally identical.

Since LEG has been extensively studied, we focus exclusively on the elemental abundances of PG1 and PG2 here. We cross-matched the data of PG1 and PG2 with APOGEE DR17 \citep{A2022}, GALAH DR4 \citep{GALAH4}, and the estimated elemental abundances from LAMOST DR10 in this study. For PG1, there are 4 common stars with LAMOST DR10, 1 common star with APOGEE DR17, and 6 common star with GALAH DR4. For PG2, the numbers are 7, 3 and 1 respectively. The limited number of overlapping samples is not surprising, as \textit{Gaia} is a full-sky survey, whereas the other datasets are obtained from ground-based surveys with more restricted sky coverage. However, the total of 11 common stars for both PG1 and PG2 still provides valuable insights into their properties.

Figure~\ref{mg} shows the [Mg/Fe] and [Al/Fe] distributions as functions of metallicity for PG1, PG2 as well as GSE stars for comparison. The PG1 common stars have wide spread of [Mg/Fe] from $-0.07$ to 0.36 dex. The relatively metal-rich ([Fe/H]$>-1$) and Mg-enhanced ([Mg/Fe]$>$0.2) PG1 common stars exhibit distributions similar to those of the high-$\alpha$ thick disk. However, there are also metal-rich PG1 stars with lower [Mg/Fe] values, which follow a distribution pattern resembling that of GSE stars. Therefore, PG1 likely represents a mixture of a low-eccentricity, metal-rich component of GSE and the thick disk. This interpretation is further supported by the low [Al/Fe] values observed in some PG1 stars, as low [Al/Fe] values are indicative of origins in dwarf galaxies \citep{Hasselquist2021,Fernandes2023}. In contrast, 9 out of the 11 PG2 common stars exhibit consistent [Mg/Fe] values, ranging between 0.3 and 0.4 dex, along with relatively high and positive [Al/Fe] values (although [Al/Fe] values are available for only 4 PG2 stars). From the current observational view, PG2 is highly likely to be an in-situ substructure, although it appears as a clump in the integral-of-motion space. This substructure is probably unrelated to Aleph, as Aleph exhibits lower $\alpha$-element enhancement, with [$\alpha$/Fe] $<$0.27. In contrast, PG1 represents a mixture of previously unidentified prograde GSE stars and thick disk stars. PG2, on the other hand, is more likely associated with the thin disk as an in-situ substructure. It is important to emphasize that the current sample size is relatively limited, and more precise determinations of the origins of these two substructures will require additional data releases in the future.

\section{Summary}\label{summary}
In this work, we apply a CNN method to map low-resolution spectra from LAMOST DR10 to APOGEE stellar labels and provide predictions for stellar parameters and elemental abundances for 1,100,858 giants. The results from the test samples are consistent with previous studies and demonstrate improved precision. For $T_{\text{eff}}$ and log $g$, the method yields uncertainties of $\sim$ 50 K and 0.13 dex with MAE values of 33.13 K and 0.09 dex. The uncertainties for metallicity ([Fe/H]) and overall metallicity (M/H) are 0.06 dex and 0.15 dex.
For $\alpha$-abundances like C, N, O, Mg, Si and Ca, the uncertainties are 0.02 dex, 0.03 dex, 0.06 dex, 0.06 dex, 0.01 dex and 0.02 dex respectively. The overall $\alpha$-abundance ([$\alpha$/M]) has an uncertainty of 0.05 dex. Most of the elemental abundances have MAE values ranging from 0.02 dex to 0.04 dex. As for MP and VMP stars, [Mg/Fe], [Si/Fe] and [$\alpha$/M] abundances still have reliable values. Our model also demonstrates superior performance compared to previous methods, such as those using astroNN \citep{Li2022} and LASP estimates, with the exception of surface gravity. For $T_{\text{eff}}$, the mean uncertainties remain below 30 K, while for $\alpha$ elements, the mean uncertainties are below 0.02 dex across the entire range of (S/N)$_g$. C and N elements exhibit slightly larger mean uncertainties ($>0.02$ dex) at low (S/N)$_g$, while $\log g$ shows mean uncertainties below 0.07 dex across the range of (S/N)$_g$. Based on the catalog, we also analyzed the distributions of the previously identified substructures PG1 and PG2 in the [Mg/Fe]-[Fe/H] and [Al/Fe]-[Fe/H] planes. Our analysis reveals that PG1 is likely a mixture of prograde GSE stars, characterized by low [Mg/Fe] and [Al/Fe] values, and the in-situ thick disk population. In contrast, PG2 is probably an in-situ substructure associated with the thin disk. Given the limited sample size, future surveys are necessary to conduct a more comprehensive analysis of PG1 and PG2. Additionally, simulations are required to provide evidence for the extent to which a GSE-like merger could produce prograde debris, further clarifying the origins and dynamics of these substructures.
The completion of the catalog will offer more insights into Galactic evolution and accretion history.

\section*{Acknowledgments}
 This work was supported by National Key R\&D Program of China No. 2024YFA1611900, and the National Natural Science Foundation of China (NSFC Nos. 11973042, 11973052). This work is based on data acquired through the Guoshoujing Telescope. Guoshoujing Telescope (the Large Sky Area Multi-Object Fiber Spectroscopic Telescope LAMOST) is a National Major Scientific
Project built by the Chinese Academy of Sciences. Funding for the project has been provided by the National Development and Reform Commission. LAMOST is operated and managed by the National Astronomical Observatories, Chinese Academy of Sciences.

Funding for the Sloan Digital Sky Survey IV has been provided by the Alfred P. Sloan Foundation, the U.S. Department of Energy Office of Science, and the Participating Institutions.

SDSS-IV acknowledges support and 
resources from the Center for High 
Performance Computing  at the 
University of Utah. The SDSS 
website is \url{www.sdss4.org}.

SDSS-IV is managed by the 
Astrophysical Research Consortium 
for the Participating Institutions 
of the SDSS Collaboration including 
the Brazilian Participation Group, 
the Carnegie Institution for Science, 
Carnegie Mellon University, Center for 
Astrophysics | Harvard \& 
Smithsonian, the Chilean Participation 
Group, the French Participation Group, 
Instituto de Astrof\'isica de 
Canarias, The Johns Hopkins 
University, Kavli Institute for the 
Physics and Mathematics of the 
Universe (IPMU) / University of 
Tokyo, the Korean Participation Group, 
Lawrence Berkeley National Laboratory, 
Leibniz Institut f\"ur Astrophysik 
Potsdam (AIP),  Max-Planck-Institut 
f\"ur Astronomie (MPIA Heidelberg), 
Max-Planck-Institut f\"ur 
Astrophysik (MPA Garching), 
Max-Planck-Institut f\"ur 
Extraterrestrische Physik (MPE), 
National Astronomical Observatories of 
China, New Mexico State University, 
New York University, University of 
Notre Dame, Observat\'ario 
Nacional / MCTI, The Ohio State 
University, Pennsylvania State 
University, Shanghai 
Astronomical Observatory, United 
Kingdom Participation Group, 
Universidad Nacional Aut\'onoma 
de M\'exico, University of Arizona, 
University of Colorado Boulder, 
University of Oxford, University of 
Portsmouth, University of Utah, 
University of Virginia, University 
of Washington, University of 
Wisconsin, Vanderbilt University, 
and Yale University.

\section*{Data Availability}
This study uses publicly available from LAMOST DR10 and APOGEE DR17.



\bibliographystyle{mnras}
\bibliography{liu_MN} 





\bsp	
\label{lastpage}
\end{document}